\documentclass[aps,prx,notitlepage,superscriptaddress,longbibliography,twocolumn,nofootinbib,floatfix]{revtex4-2} 

\usepackage[utf8]{inputenc}
\usepackage{graphicx}
\usepackage{hyperref}
\usepackage[dvipsnames]{xcolor}

\usepackage[compat=1.0.0]{tikz-feynman}

\usepackage{lipsum}

\usepackage{amsmath,amsthm,amssymb,mathtools}
\usepackage{multirow}
\usepackage{braket}
\usepackage{bbm,bm}
\usepackage[bb=boondox]{mathalfa}
\usepackage{dsfont}

\usepackage{iftex}

\DeclareFontFamily{U}{matha}{\hyphenchar\font45}
\DeclareFontShape{U}{matha}{m}{n}{
      <5> <6> <7> <8> <9> <10> gen * matha
      <10.95> matha10 <12> <14.4> <17.28> <20.74> <24.88> matha12
      }{}
\DeclareSymbolFont{matha}{U}{matha}{m}{n}
\DeclareMathSymbol{\muparrow}{3}{matha}{"D2}
\DeclareMathSymbol{\mdownarrow}{3}{matha}{"D3}
\DeclareMathSymbol{\mupdownarrow}{3}{matha}{"D9}

\DeclareFontFamily{U}{mathb}{\hyphenchar\font45}
\DeclareFontShape{U}{mathb}{m}{n}{
      <5> <6> <7> <8> <9> <10> gen * mathb
      <10.95> mathb10 <12> <14.4> <17.28> <20.74> <24.88> mathb12
      }{}
\DeclareSymbolFont{mathb}{U}{mathb}{m}{n}
\DeclareMathSymbol{\mdownuparrows}{3}{mathb}{"D7}

\usepackage{makecell}

\usepackage{float}

\definecolor{dgreen}{rgb}{0,0.666,0}
\definecolor{dorange}{rgb}{0.666,.333,0}

\newcommand{\pd}{{\phantom{\dagger}}}

\makeatletter
\pgfmathdeclarefunction{erf}{1}{%
  \begingroup
    \pgfmathparse{#1 > 0 ? 1 : -1}%
    \edef\sign{\pgfmathresult}%
    \pgfmathparse{abs(#1)}%
    \edef\x{\pgfmathresult}%
    \pgfmathparse{1/(1+0.3275911*\x)}%
    \edef\t{\pgfmathresult}%
    \pgfmathparse{%
      1 - (((((1.061405429*\t -1.453152027)*\t) + 1.421413741)*\t 
      -0.284496736)*\t + 0.254829592)*\t*exp(-(\x*\x))}%
    \edef\y{\pgfmathresult}%
    \pgfmathparse{(\sign)*\y}%
    \pgfmath@smuggleone\pgfmathresult%
  \endgroup
}
\makeatother

\begin{document}
\title{Projective Measurements: Topological Quantum Computing\\
with an Arbitrary Number of Qubits}

\author{Themba Hodge}
\affiliation{School of Physics, The University of Melbourne, Parkville, VIC 3010, Australia}


\author{Philipp Frey}
\affiliation{School of Physics, The University of Melbourne, Parkville, VIC 3010, Australia}

\author{Stephan Rachel}
\affiliation{School of Physics, The University of Melbourne, Parkville, VIC 3010, Australia}

\begin{abstract}
Topological quantum computing promises intrinsic fault tolerance by encoding quantum information in non-Abelian anyons, where quantum gates are implemented via braiding. 
While braiding operations are robust against local perturbations, a critical yet often overlooked challenge arises when scaling beyond two qubits: the naive extension of braiding-based gates fails to support even the full Clifford group. 
To overcome this limitation, we incorporate projective measurements that enable transitions between different qubit encodings, thus restoring computational universality.
We perform many-body simulations of braiding dynamics augmented with measurement-based switching, explicitly preparing the Bell state and GHZ state for systems of two and five qubits, respectively. 
Furthermore, we execute a random unitary circuit on five qubits, achieving a fidelity exceeding 99\%. 
We analyze the circuit’s robustness by studying its fidelity dependence on total braid duration and static potential disorder. 
Our results show that the fidelity remains above 99\% for moderate disorder, underscoring the intrinsic fault tolerance of the architecture. 
Finally, we demonstrate a random circuit on a ten-qubit system to showcase the scalability of our techniques.

\end{abstract}

\maketitle








{\it Introduction.---}
The advantage of topological quantum computation (TQC) lies in the implementation of quantum gates via braiding. The pairwise exchange of Majorana zero modes (MZMs) leads to an action on the degenerate many-body ground state manifold that is identical to the action of a unitary operator on the Hilbert space that it spans\,\cite{nayak-96npb529,Ivanov2001,georgiev-06pra235112,Nayak2008,Bonderson2011,Barkeshli2013,beenakker20}. 
In fact, for up to two qubits, all Clifford gates can be implemented via braiding\,\cite{lahtinen-17spp021} and Majorana hybridization, i.e., a controlled energy splitting induced by proximity\,\cite{Bonderson2010,hodge2025}. 
Crucially, this implementation of quantum gates is inherently fault-tolerant due to topological protection against local noise.



A major obstacle to the practical realization of TQC concerns the encoding of qubits using Majorana operators in a way that supports universal computation. In 1D topological superconductors, two end-localized Majoranas define a non-local fermionic mode\,\cite{kitaev2001}, with the vacuum and occupied states typically mapped to $\ket{0}$ and $\ket{1}$, respectively. However, qubit counting is nontrivial: $2N$ Majoranas do not yield $N$ independent qubits due to global fermion parity constraints, which reduce the accessible Hilbert space by half. This conservation law also blocks coherent superpositions of different parity sectors via braiding alone, limiting the ability to dynamically prepare arbitrary quantum states.

Any solution to this problem involves the addition of auxiliary qubits. 
This may be done in several different ways, each of which constitutes a particular \textit{encoding}.
However, it can be shown that no one encoding alone allows for universality if only braiding and hybridization between two Majoranas is considered. 
In fact, universality may not be achieved through braiding without the addition of hybridization, but the full set of Clifford operations may be achieved if we combine at least two different encodings and the ability to map between them via projective measurements.

Here, we utilize the \textit{sparse} and \textit{dense} encodings\,\cite{dassarma-15npjqi15001}. 
The sparse encoding assigns four Majoranas to each logical qubit, meaning that each qubit comes with its own ancillary, and allows for the implementation of all single-qubit Clifford gates via braiding of the four Majoranas. 
In addition, the famous magic gate may be implemented via hybridization.
The downside of this encoding is that different qubits cannot be entangled with one another. 
The dense encoding, on the other hand, assigns $2N+2$ Majoranas to $N$ logical qubits and therefore requires only one ancillary. 
It readily allows for mutual entanglement. 
However, the full set of single-qubit Clifford gates is available only on the first and last logical qubit.

The limitation of each encoding can be overcome by mapping between them using projective measurements. 
This idea is implicit in \cite{Bravyi2006,Nayak2008} and a few other publications since, but is rarely, if ever, spelled out in detail. 
We present an accompanying in-depth exposé\,\cite{Frey2025}, along with the necessary theoretical background. 
This includes details for our time-dependent Pfaffian method\,\cite{mascot2023}, an application of Hartree--Fock--Boguliubov theory\,\cite{ring1980,bertsch2012,carlsson2021} for dynamically simulating universal TQC.

In this Letter, we briefly introduce the method and relevant operators required to perform the projective calculations necessary to map between the sparse and dense encoding. 
We then utilize this method to first demonstrate the dynamic preparation of the Bell state, and then expand this to prepare the maximally entangled GHZ state on $N=5$ qubits.
We emphasize both the utility, along with the universality of the method, by the dynamic execution of a random unitary circuit on five qubits, matching the exact unitary with fidelity beyond 99\%. 
Next, we explore the dependence of this fidelity on the total braid time $T_{\rm braid}$ and static potential disorder $V$. 
The resulting fidelity plot features extended areas of $>$99\%, which shows that TQC is indeed protected against disorder. 
Finally, we push the computational boundary and reach a circuit size of $N=10$ qubits, corresponding to 40 MZMs.



\begin{figure}[t!]
    \centering
    \includegraphics[scale=0.95]{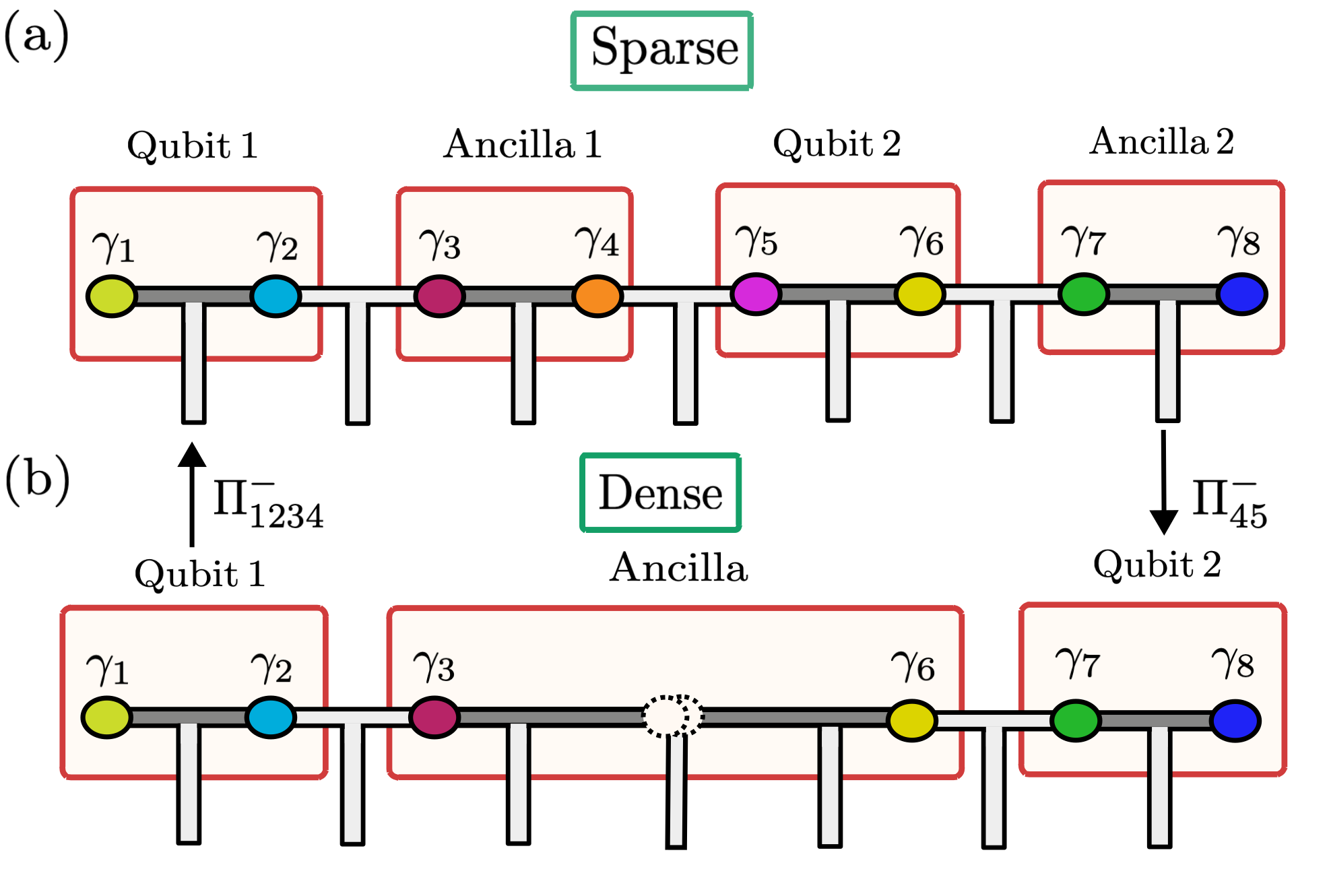}
    \caption{Logical qubit encodings on T-junction geometry. 
    (a) Sparse encoding for two qubits, each consisting of a physical and a parity-conserving ancilla qubit. 
    (b) Dense encoding for two qubits, consisting of physical qubits and a single ancilla qubit.} 
    \label{fig:Fig1}
\end{figure}

{\it Method.---}
All simulations in this paper will be conducted on the Kitaev chain \cite{kitaev2001}, the minimal model required to host MZMs. 
The time-dependent Hamiltonian is given by
\begin{equation}
\begin{aligned}
H(t) = -&\sum_a \mu_a(t) c_a^\dagger c_a^\pd
- \tilde{t} \sum_a \left(
    c_a^\dagger c_{a+1}^\pd + \rm{h.c.}
\right)\\
&+ \sum_a \left(
    \Delta_p c_a^\dagger c_{a+1}^\dagger + \rm{h.c.}
\right),
\end{aligned}
\end{equation}
where $\mu_a(t)$, $\tilde{t}$ and $\Delta_p$ correspond to a time-dependent potential term, tunneling strength, and $p$-wave superconducting pairing strength. 
Note, $\Delta_p=e^{i\phi}|\Delta_p|$, where $\phi\in[0,2\pi]$.
The topological regime, characterized by MZMs on the boundary, is specified when $|\mu|<2\tilde{t}$. 
In this way, we may ramp the chemical potential locally between the topological and trivial regimes to spatially move the MZMs along the chain. 
By joining three chains in the form of three legs, with the pairing phase in the horizontal (vertical) legs $\phi=0$ ($\phi=\frac{\pi}{2}$), we are able to form a T-junction \cite{alicea2011}, the minimal structure required to braid two MZMs.
A schematic of the structure for two qubits is given in Fig.\,\ref{fig:Fig1}.
$2n-1$ such T-junctions may then be joined together to construct a model which can braid $2n$ MZMs, with a single leg in between each neighboring set of MZMs. 
\begin{figure*}[t!]
    \centering
    \includegraphics{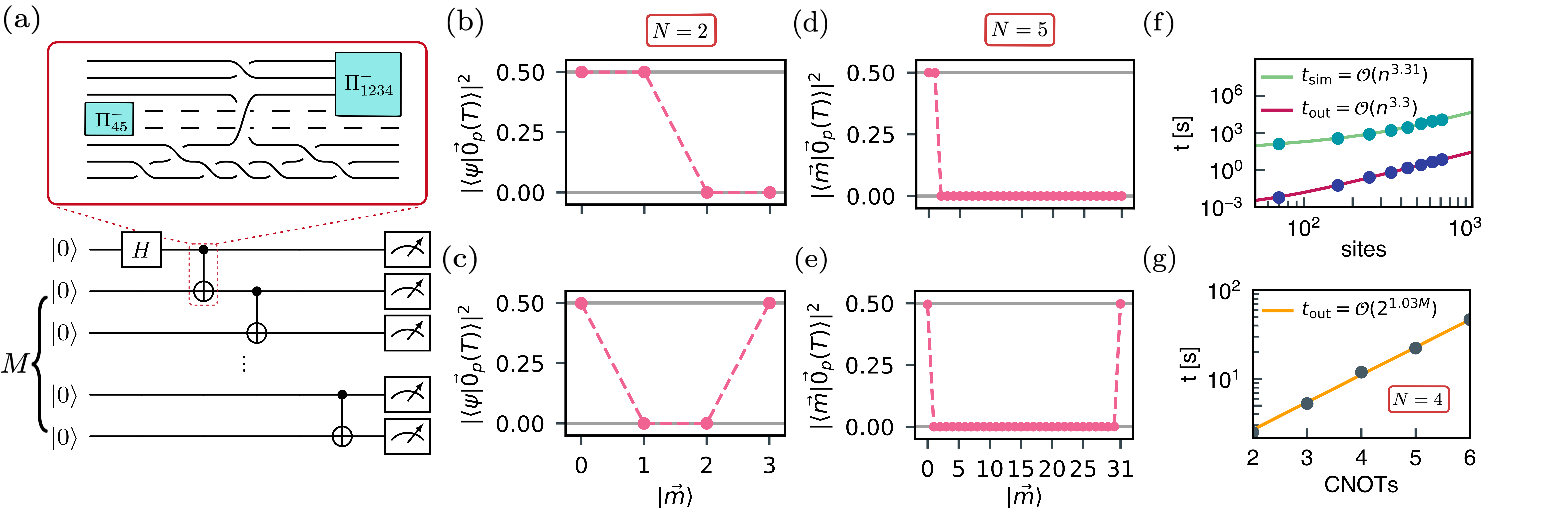}
    \caption{N-qubit GHZ state preparation. (a) Circuit diagram of GHZ state.
    (b, c) Transition probabilities, $|T_{\vec{m}\vec{0}}(T)|^2$, of all qubit states for a two-qubit system, after a Hadamard on the first qubit (b) and a subsequent CNOT gate (c). Qubit states are labeled as binaries, see main text. 
    (d, e) Same as (b, c) but for $N=5$ GHZ state.
    (f) Average total readout time of a single overlap, $T_{\vec{0}\vec{0}}(T)$, after a single H gate on an $N$ sparse qubit system.
    (g) Same as (f) but after $M$ CNOTs for $N=4$. Solid lines in (f, g) are fits as indicated by the legend.
    Parameters used are $(\tilde{t},\,\mu,\,\mu_{\rm{triv}},\,\Delta,\,L)=(1,0.02,10.4,0.98,7a_0)$ for all simulations, with $T_{\rm braid}=2400\hbar/\tilde{t}$ for (b-e) and $T_{\rm braid}=1500\hbar/\tilde{t}$ for (f, g).}   
    \label{fig:Fig2}
\end{figure*}

As discussed in the Introduction, and in detail in \cite{Frey2025}, for the purpose of scalable quantum computation, projective measurements between sparse and dense encodings are a viable pathway towards the implementation of a universal gate set.
A schematic of both is shown in Fig.\,\ref{fig:Fig1}.
In essence, this method relies on the projection of a logical two-qubit state in the sparse encoding, $|\vec{n}\rangle_{s}$, into the dense encoding $|\vec{n}\rangle_d$. 
For simplicity and in line with the discussion in \cite{Frey2025}, consider a set of Majorana bound states initialized in the $d_{2i-1,2i}=\frac{1}{2}\big(\gamma_{2i-1}+i\gamma_{2i}\big)$ basis. 
A set of projection operators encoding the projective measurements are given by
\begin{equation}
    \begin{aligned}
     &\Pi_{45}^{-}=d_{45}d^{\dag}_{45},\quad \Pi_{1234}^{-}=\frac{1}{2}\left(1 + Q_{s,1} \right)
    \end{aligned}
\end{equation}
where $Q_{s,1}=-\gamma_1\gamma_2\gamma_3\gamma_4$.
Here, $\Pi^-_{45}$ will project the state into the subspace where qubit 1 and 2 have even joint parity. 
This projection is made possible via MZM fusion of $\gamma_4$ and $\gamma_5$, which will measure the parity sector of the $-i\gamma_4\gamma_5$ bound state. 
Subsequently, $\Pi^-_{1234}$ projects the state back into the sparse encoding of total even parity. 
\begin{equation}
 \Pi_{45}^{-}|\vec{n}\rangle_{s}=\frac{1}{\sqrt{2}}|\vec{n}\rangle_d, \quad \Pi_{1234}^{-}|\vec{n}\rangle_d=\frac{1}{\sqrt{2}}|\vec{n}\rangle_s.
\end{equation}
We stress that we may equivalently project into the subspace of odd joint parity subspace, without any loss of quantum information. 
The advantages of utilizing this process is immediately evident. 
While all single qubit gates may be safely done in the sparse encoding, we project into the dense encoding in order to enact any string of CZ and/or CNOT gates between two neighboring qubits, i.e., to entangle the qubits. 
We then project back onto the sparse encoding once we are finished with these two-qubit processes. 

For the purpose of classical simulation, 
we read the quantum information by finding the transition matrix, $T_{\vec{m}\vec{n}}(t)$, given by 
\begin{equation}
\begin{aligned}
    T_{\vec{m}\vec{n}}(t)&=2
    \langle\vec{m}|U(t,t_{2})\Pi^-_{1234}U(t_{2},t_{1})\Pi^-_{45}U(t_{1},t_{0})|\vec{n}\rangle\\
    &=2\braket{\vec{m}|\vec{n}_p(t)}\label{eq:transitionmat}
\end{aligned}
\end{equation}
where the time-evolution operator, $U(t_i,t_j)$ is given by $U(t_i,t_j)=\mathcal{T}\exp\big(-i\int^{t_i}_{t_j}H(t')dt'\big)$. 
Here $U(t,t_{2})$ and $U(t_1,t_{0})$ correspond to the time evolution associated with single-qubit processes in sparse encoding, and $U(t_{2},t_{1})$ with two-qubit gates in dense encoding, with the normalization factor $\sqrt{2}$ required to conserve total probability.
We denote the state $\ket{\vec{n}_p(t)}$ as the \emph{projected state}, corresponding to the quantum state recovered after the application of both single and two-qubit unitary time-evolutions, along with the associated projective measurements required to enact these gates.
As such, by calculating $T_{\vec{m}\vec{n}}(T)$, with $T$ the total simulation time, we recover all quantum information of the final state $\ket{\vec{n}_p(T)}$.
{\it Results.---}
For the remainder of this Letter, we will demonstrate the efficacy of utilizing this projection method to classically simulate a scalable quantum computer on MZM-based platforms. 
Each gate in this paper is performed dynamically, either through a set of braids ($\sqrt{X}$, $S$, CNOT) or through a timed MZM hybridization routine ($T$)\,\cite{hodge2025}.
A schematic of the braids required for a CNOT gate in the dense encoding, in the basis discussed in the Method section, is given schematically in Fig.\,\ref{fig:Fig2} (a).
In addition, all time-dependent overlaps are calculated using the \emph{time-dependent Pfaffian method}\,\cite{mascot2023}, an exact method for finding arbitrary overlaps of many-body states in a free theory.  
Finally, all computational results were computed in the $X$-basis \cite{Nayak2008}, where for $k\in\{1,..,N\}$, the computational Hilbert-space is stabilized by the pairs of operators $-i\gamma_{4k-2}\gamma_{4k-1}$, $-i\gamma_{4k-3}\gamma_{4k}$. 
Here, $\gamma_{4k-3}$ and $\gamma_{4k+4}$ projected out when moving to dense encoding, retaining $\gamma_{4k-2},.., \gamma_{4k+2}$ in the dense encoding.



{\it GHZ State.---}
We begin with two qubits.
First, we institute an H gate on the first qubit, mapping $|00\rangle \to \frac{1}{\sqrt{2}}\big(|00\rangle+|10\rangle\big)$, as demonstrated in Figs.\,\ref{fig:Fig2} (b, c).
Using the projective method detailed in the previous section, we map the quantum information from the sparse encoding to the dense encoding, where we can perform a CNOT gate, which maps $|10\rangle\to|11\rangle$, keeping $|00\rangle$ unaffected. 
This is shown in Fig.\,\ref{fig:Fig2} (b), (c), which demonstrates a probability transfer from the $\ket{10}$ state to the $\ket{11}$ state, thus corresponding to the successful implementation of the projective method. 
We note that all logical states are indexed using a binary mapping $f$: $\mathcal{H}\to \mathbb{Z}$, where $f(|n_1,n_2...,n_N\rangle)=\sum^{N-1}_{i=0}2^in_{i+1}$ in the sparse encoding.

We then emphasize the scalability of the method by introducing the generalized GHZ state on $N$ sparse qubits, with the definition $\ket{\rm{GHZ}(N)}=\frac{1}{\sqrt{2}}\big(|0\rangle^{\otimes_N}+\ket{1}^{\otimes_N}\big)$, where, for an arbitrary state $\ket{a}$, $\ket{a}^{\otimes_M}=\bigotimes^M_{i=1} \ket{a}$. 
The corresponding circuit diagram required to produce this state from the initial $\ket{\vec{0}}$ state is given in Fig.\,\ref{fig:Fig2} (a).
 After the first H gate, $\ket{\vec{0}} \to \frac{1}{\sqrt{2}}\big(\ket{0}+\ket{1}\big)\otimes\ket{0}^{\otimes _{N=4}}$, as shown in Fig.\,\ref{fig:Fig2} (d). 
A string of CNOT gates is then implemented on each pair of neighboring sparse qubits, as indicated schematically in Fig.\,\ref{fig:Fig2} (a).  
This maps the state to the generalized GHZ state on five qubits, defined by $\ket{\vec{0}_p(T)}=\frac{1}{\sqrt{2}}\big(\ket{0}^{\otimes_{N=5}}+\ket{1}^{\otimes_{N=5}}\big)=\ket{\rm {GHZ}(5)}$, as demonstrated in Fig.\,\ref{fig:Fig2} (e).

Next, we investigate the computational scaling of the classical simulation method.
This is highlighted in Fig.\,\ref{fig:Fig1} (f, g), where we track the average readout time as a function of $N$ after a single-qubit gate (f), and then investigate how the readout time scales as a function of the projective operations (g). 
In Fig.\,\ref{fig:Fig1} (f), we apply a single $H$-gate on a system with $N$ sparse qubits, with both sets of data fit against the number of lattice sites $n\equiv n(N)=70+92(N-1)$.
First, we consider the time required to construct the time-evolution operator $U(T,0)$. 
Here, we find a polynomial scaling in the simulation time, with $t_{\rm{sim}}=\big[(0.024\,n)^{3.31}+1.82\,n\big]\,\, \rm s$ for $N\in[1,9]$ after 50 simulations.
We extend this analysis to the readout time of a single overlap, $T_{\vec{0}\vec{0}}(T)$. 
First, we note that the contraction matrix $M_{\vec{0}\vec{0}}(t)$, where $T_{\vec{0}\vec{0}}(T)\propto \rm pf[M_{\vec{0}\vec{0}}(T)]$, has size $2n-p$, with $p$ corresponding to the size of the empty sector after the Bloch-Messiah decomposition (see \cite{Frey2025} for further details). 
Assuming the empty sector to be small vs.\ the size of the system (i.e., $2n\gg p$), which is consistent with our simulations, we fit the average time taken against $n$, thus providing a prediction for the readout time vs.\ the physical parameter set utilized. 
Here, we find the readout time scales as $t_{\rm{read}}=\big[(2.5\times 10^{-3}\,n)^{3.30}+4.6\times 10^{-5}\,n\big]\,\,\rm s \sim \mathcal{O}(n^{3.3})$ in our data set, approaching the expected $\mathcal{O}(n^3)$ scaling of the Pfaffian calculation\,\cite{Wimmer2012}, which dominates the routine in the large $n$ regime.
However, setting $N=4$ and replacing the Hadamard gate with a string of CNOTs on the first qubit, we highlight the exponential increase in computational time with every projective operation.
As shown in Fig.\,\ref{fig:Fig1} (g), 
we find the readout time scales as $t_{\rm{read}}=\big[0.64\times 2^{1.03M}\big] \,\,\rm s$, in line with the increase in the number of contractions required to find $T_{\vec{m}\vec{n}}(T)$, which grows as $2^M$. 
\begin{figure}[t!]
    \centering
    \includegraphics{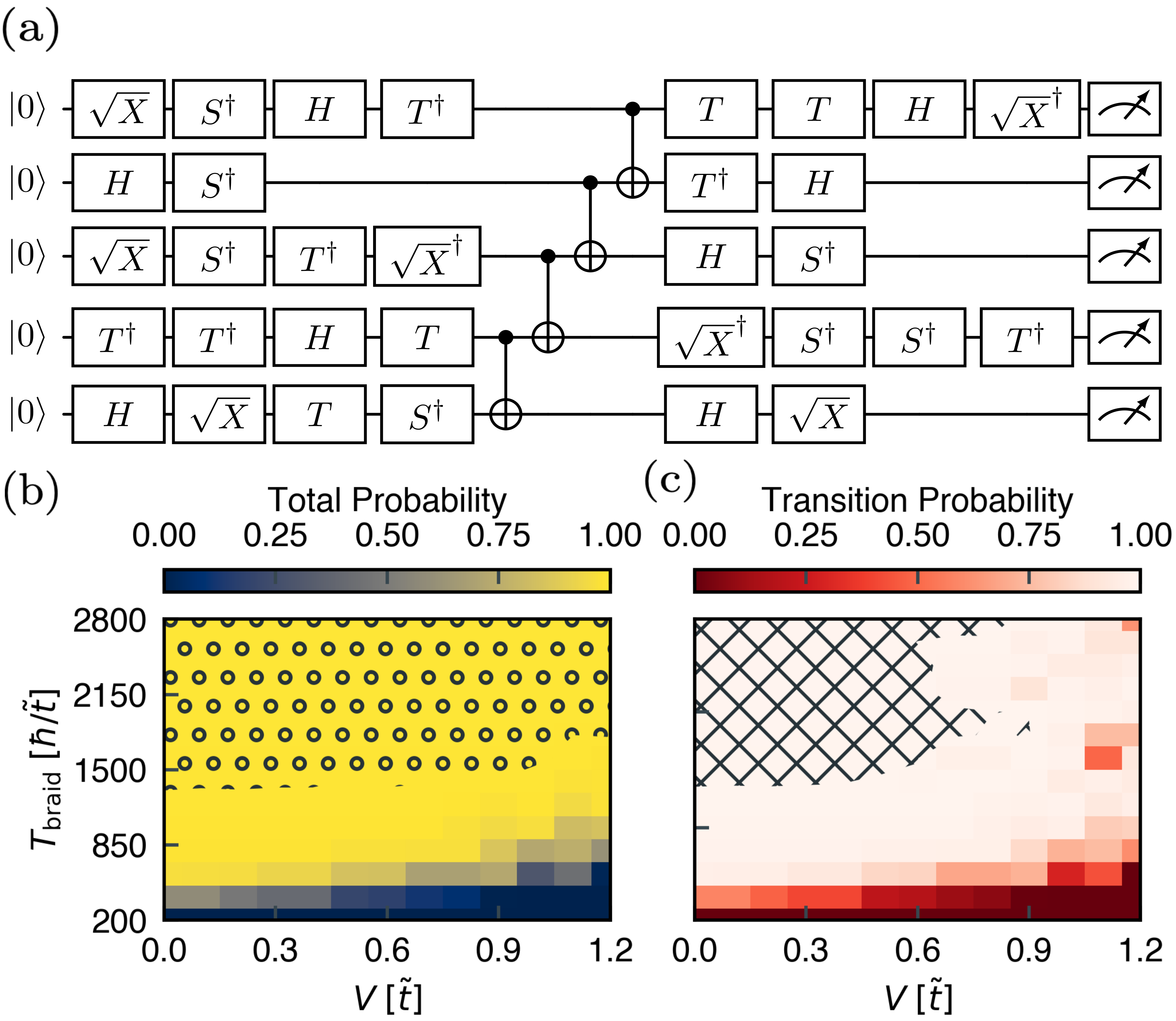}
    \caption{Random unitary circuit in the presence of static disorder. 
    (a) Schematic of the random unitary circuit.
    (b) Total probability within the MZM subspace, $\sum_n |T_{\vec{n}\vec{0}}(T)|^2$, between the time-evolved initial state $|\vec{0}(T)\rangle$ and the target state $|t\rangle$ (c), with regions of probability$\,>\,99\%$ highlighted by hatches overlaying the data. (c) Same as (b) for the transition probability, i.e., fidelity. In (b, c) quantities are plotted as a function of braid time $T_{\rm{braid}}$ and disorder strength $V$. 
    Parameters used in all simulations are $(\tilde{t},\,\mu,\,\mu_{\rm{triv}},\,\Delta,\,L)=(1,\,0.02,\,8.8,\,0.9,\,8a_0).$}
    \label{fig:Fig3}
\end{figure}

While we observe exponential scaling in the projective measurements, we highlight two vital points: 
First, while the theoretical framework for the projective method is laid out in \cite{Frey2025}, here, we demonstrate how it is a feasible framework to perform large-scale braiding simulations for systems with $N$ qubits.
Second, each overlap in the output routine may, in essence, be calculated sequentially.  
This removes the cost of storing the full quantum state throughout the process, which will lead to an exponential scaling in data storage with the number of qubits in the system.
Thus, the number of overlaps done in parallel may be optimized against both total CPU time and memory constraints, providing a malleable procedure to extract quantum information.

                                 



{\it Random Circuit.---}
We now look to demonstrate the feasibility of the method to simulate any generic quantum computation.  
In this section, we will focus on the simulation of a random circuit, enacted on an $N=5$ sparse qubit system, with the circuit generated given in Fig.\,\ref{fig:Fig3} (a).
Further, we also consider the effects of static disorder on the system, thus emphasizing the utility of this method to probe the physical conditions of the underlying platform hosting MZMs. 
 We do this by, at each point $l$ on the adjoined $T$-junction, adding an impurity term to the Hamiltonian of the form $w_lc_{l}^\dagger c_{l}^{\phantom{\dagger}}$.
Here, each onsite $w_l$ is sampled from a uniform distribution spanning from $[-V,V]$, thus establishing a static disorder configuration which shifts the local energetics of the lattice structure. 
Each data point in Fig.\,\ref{fig:Fig3} corresponds to a different effective disorder configuration, thereby providing a picture for the ensemble effect of the perturbation on the physics of large scale MZM braiding.
First, while disorder will perturb the energy landscape, thus affecting the timed hybridization routine necessary to implement the $T$-gate, we overcome this issue by calibrating for the required time to implement the $T$-gate on each Majorana bound-state, before the implementation of the circuit, thus, circumventing this issue. 
Next, as discussed in \cite{Peeters2024}, for braiding, static disorder introduces two major error sources: \emph{diabatic error} as the effective bulk gap is reduced in large $V$ regimes, leading to an increased likelihood of loss of quantum information to the bulk, along with the potential for increased \emph{hybridization error}, due to local impurities driving MZM wavefunction broadening. 
This leads to an increased likelihood of wavefunction overlap with other MZMs in the system, and thus, unwanted $\rm{SO}(4N)$ rotations of the MZM operators \cite{cheng2011,Frey2025}. 
We reveal the effects of adiabatic error in Fig.\,\ref{fig:Fig3} (b), where the total probability measured in the ground state manifold approaches 1 as the time of each braid $T_{\rm braid}$ approaches $2800\frac{\hbar}{\tilde{t}}$.
This confirms the entering of the adiabatic regime in the large limit of $T_{\rm braid}$, whereby the total state probability remains conserved with the ground state manifold. 
However, as shown in Fig.\,\ref{fig:Fig3} (c), while the system is clearly in the adiabatic regime, as $V$ increases, the system accrues error within the MZM subspace.
For lower values of $V$, we see that these hybridization effects are diminished, leading to a stable region where the fidelity $|\langle t|\vec{0}_p(T)\rangle|^2>0.99$, regardless of the disorder configuration. 
In the case of large disorder strengths, we clearly demonstrate that these error sources remain detrimental to braiding, with no simulation satisfying a fidelity larger than $99\%$ for disorder configurations with $V\geq\tilde{t}$.
 \begin{figure}[t!]
    \centering
    \includegraphics{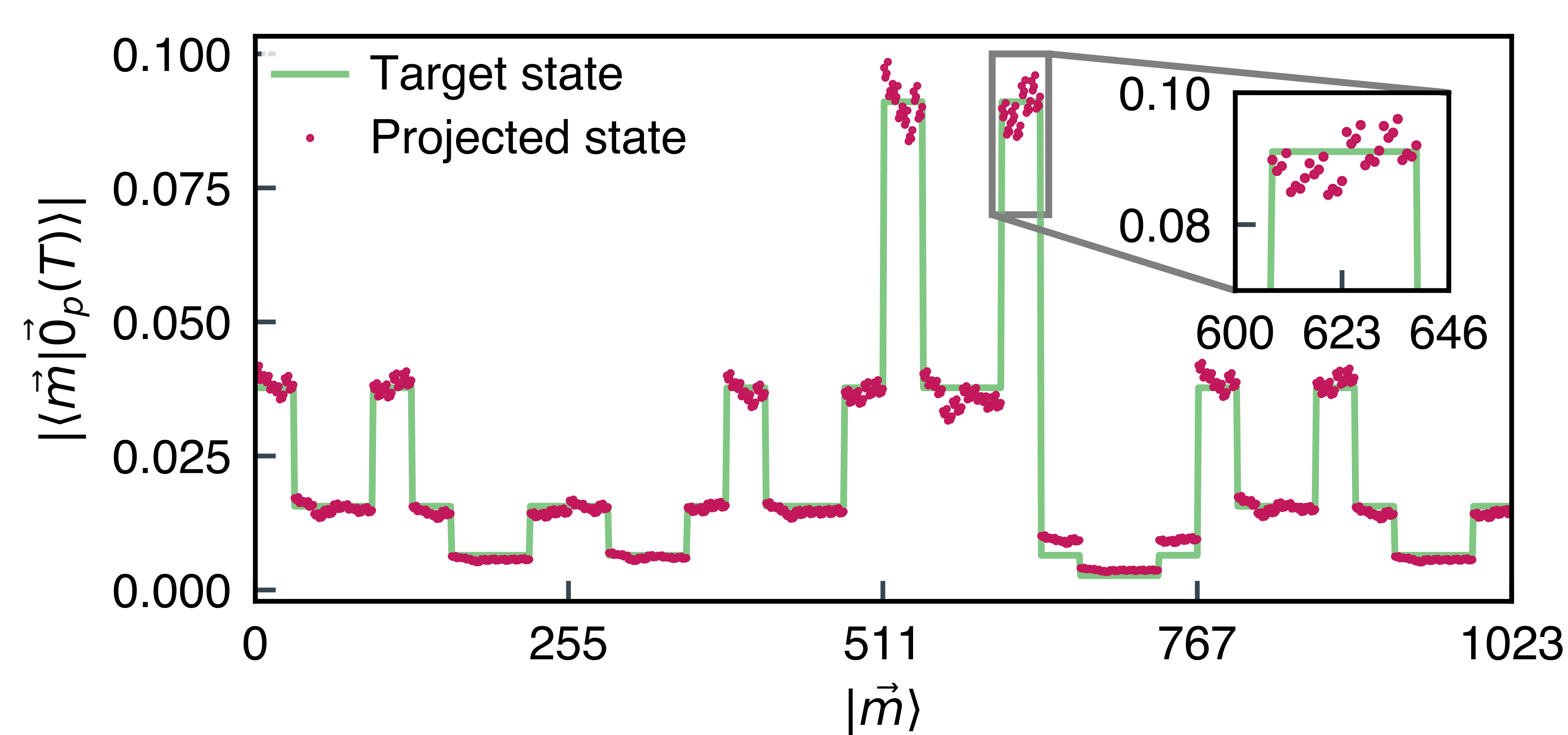}
    \caption{Transition probabilities for 10 qubit simulation of a random unitary. 
    Parameters used in this simulations are $(\tilde{t},\,\mu,\,\mu_{\rm{triv}},\,\Delta,\,L,T_{\rm braid})=(1,\,0.02,\,10.4,\,0.98,\,7a_0,2000\hbar/\tilde{t})$.}
    \label{fig:Fig4}
\end{figure}
 In essence, we have demonstrated both the scalability of the method to large circuits, whilst stressing the utility of the method to investigate the effects of various error sources on large quantum circuits.
 
 

                                 


{\it Ten-Qubit Simulation.---}
Finally, we extend this by simulating a random unitary circuit on a large-scale ten-qubit system, where we again perform a CNOT gate on each neighboring pair of qubits, sandwiched in between a randomly chosen set of single-qubit gates (see Supplement\,\cite{Supplement} for the schematics of the circuit).
In total, it contains 77 qubit gates enacted on the circuit, with 18 projective measurements, corresponding to $M=9$. 

We consider the matrix elements of the projected overlap $T_{\vec{i}\vec{0}}(T)$. 
This is shown in Fig.\,\ref{fig:Fig4}, where the absolute values of the matrix elements are plotted against the magnitudes of the target state associated with this random circuit. 
Moreover, while Fig.\,\ref{fig:Fig4} removes phase information, providing an incomplete picture of the distance between the projected and final state, we see that the fidelity, while dipping below the threshold fidelity of $99\%$, as discussed in Fig.\,\ref{fig:Fig3}, remains significant, with $|\braket{t|\vec{0}_p(T)}|^2=0.974$, with the total probability of the time-evolved state being $\sum_i|T_{\vec{i}\vec{0}}(T)|^2=0.981$. 
This suggests that while probability is lost as a result of diabatic and hybridization error, marginally deviating the result away from the target state, the non-Abelian braiding statistics is preserved over the course of the simulation.
As such, this stands as a successful simulation of a 10-qubit quantum circuit on an MZM platform, 
and demonstrative of the largest simulation of any topological quantum circuit on an MZM-based platform.

                                 


{\it Discussion and Outlook.---}
We have shown that projective measurements not only stand as a potential pathway towards universal topological quantum computation, but by using the scalable method introduced in \cite{mascot2023}, this method opens the door for the classical simulation of large-scale quantum circuits within this particular paradigm.

Although all results have been found using the spinless Kitaev nanowire, we stress that the method is fundamentally platform independent. 
As such, there is scope for future investigation on superconductor-semiconductor heterostructures\,\cite{Karzig2019,Pan2020,microsoft23,microsoft25}, magnetic superconductor hybrid systems\,\cite{nadjperge13,Ruby2015,pawlak2016,bedow2023,rachel-25pr1,hodge25fusion}, and two-dimensional higher-order platforms \cite{Zhang20203,hodge25Alter}, all of which provide scalable platforms for MZM braiding, and are eminently realizable experimentally. 
While we have chosen beneficial parameter sets to demonstrate the recovery of the non-Abelian statistical properties of the MZMs, as shown in Fig.\,\ref{fig:Fig3}, this method opens the door to investigate real-world conditions and various sources of error on quantum circuits.
This includes the effects of Pauli-qubit error\,\cite{cheng2011,hodge2025} in the MZM manifold, along with the effects of diabatic error\,\cite{Nayak2008,karzig2013,knapp2016,karzig-21prl057702}, especially in the presence of disorder\,\cite{Brouwer2011,Truong23,Pan24}.
Finally, we stress that by calculating each overlap sequentially we avoid the cost of storing the full quantum state over the course of the readout, which grows exponentially with the number of qubits. 
Thus, we have detailed an accessible procedure to
perform large-scale simulations.



\begin{acknowledgements}
    SR acknowledges discussions with Sankar das Sarma, Bela Bauer and Cole Peeters. The authors are grateful to Eric Mascot for closely related collaborations. This work was supported by the Australian Research Council (ARC) through Grants No.\ DP200101118 and DP240100168.
This research was undertaken using resources from the National Computational Infrastructure (NCI Australia), an NCRIS enabled capability supported by the Australian Government.
\end{acknowledgements}

\bibliography{bibliography/short}



\end{document}


\title{Supplementary Material\\[5pt]
Projective Measurements: Topological Quantum Computing\\
with an Arbitrary Number of Qubits}

\author{Themba Hodge}
\author{Philipp Frey}
\author{Stephan Rachel}

\affiliation{School of Physics, University of Melbourne, Parkville, VIC 3010, Australia}
\maketitle

\begin{figure}[h!]
    \centering
    \includegraphics{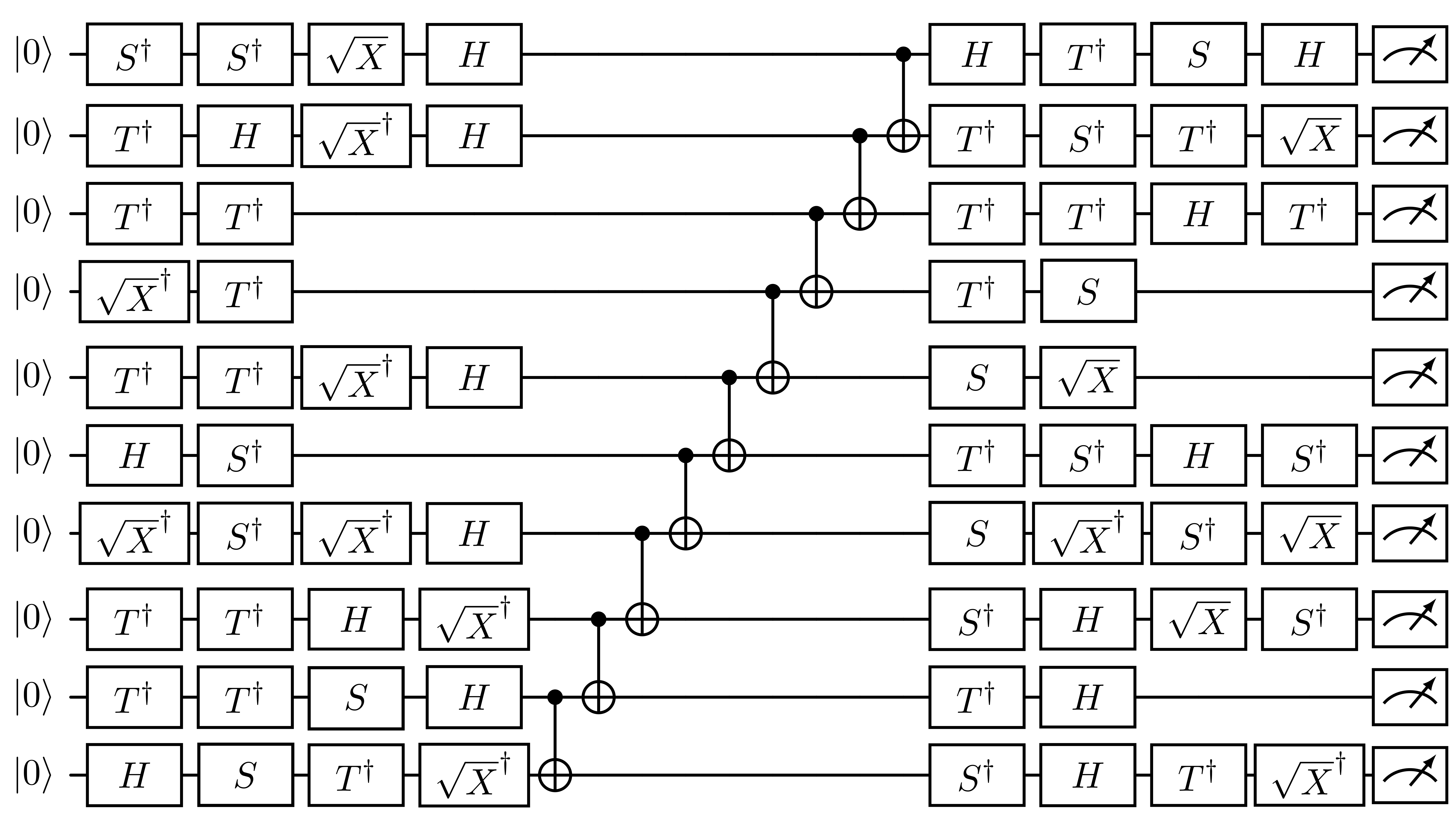}
    \caption{Schematic of the quantum circuit utilized for the ten qubit simulation in Fig.\,4 of the main text.}
    \label{fig:suppfig1}
\end{figure}
Fig.\,\ref{fig:suppfig1} provides a schematic of the quantum circuit utilized for the simulation of a ten-qubit system in Fig.\,4 of the main text, where we present the matrix elements of the projected overlap matrix $T_{\vec{i}\vec{0}}(T)$, corresponding to the quantum information retained within the logical subspace after time evolution.
\bibliography{bibliography/short}